# Modulation instability induced by periodic power variation in soliton fiber ring lasers

Zhi-Chao Luo,[1,*] Wen-Cheng Xu,[1] Chuang-Xing Song,[1] Ai-Ping Luo[1] and Wei-Cheng Chen[2]

*1. Laboratory of Photonic Information Technology, School of information and Optoelectronic Science and Engineering, South China Normal University, Guangzhou, Guangdong 510006, China*

*2. Department of Physics, Foshan University, Foshan, Guangdong 528000, China*

*Corresponding author:* zcluo@scnu.edu.cn

**Abstract**: Modulation instability with subsideband generation induced by periodic power variation in soliton fiber ring lasers is reported. We found that different wavelength shifts of subsideband generation are related to different periodic power variation. The period of power variation and wavelength shifts of subsideband can be changed by altering the linear cavity phase delay. It is also found that the periodic power variation is caused by the interaction between the nonuniform polarization state of the circulating light and the polarizer in the laser cavity.

# 1. Introduction

Modulation instability as a well-known nonlinear phenomenon has attracted considerable attention [1-8]. It can be found in many fields of physics, such as nonlinear optics, plasma physics and fluid dynamics. This phenomenon is characterized by an exponential growth of weak optical perturbations from the steady state. Modulation instability in optical fibers [1] was originally studied by Hasegawa et al. the and their theoretical prediction was observed experimentally by Tai et al. [2]. Sideband generation is a general characteristic of soliton fiber laser whose optical spectrum exhibits sharp, discrete spectral components. Although modulation instability was first deemed as the origin of sideband generation [9], it was later found that sideband generation was caused by the constructive interference between soliton pulse and dispersive waves [10,11]. The soliton pulse circulating in the ring cavity periodically endures the perturbations such as the gain provided by erbium-doped fiber and output loss. Therefore, after each perturbation, soliton pulses shed any excess energy into the so-called dispersive waves due to the reshape of the soliton pulses. As the dispersive waves are continuous waves (CW), they are unstable for the modulation instability if their strength is strong enough. The subsideband generation in a fiber ring laser was first studied by Tang et al. [12]. The subsidebands are the discrete spectral components beside the sidebands on the optical spectrum. They have shown that the generation of subsideband was a result of modulation instability of dispersive waves induced by the periodic dispersion variation in the ring cavity. However, in a soliton fiber ring laser both periodic dispersion variation [13] and periodic power variation [14] experienced by the dispersive waves can result in the modulation instability. Although in a fiber ring laser the dispersive waves periodically experience gain provided by the erbium-doped fiber and loss at the output coupler which lead to periodic power fluctuation, this kind of power variation is too small and the modulation instability with subsideband generation can not be observed in the experiment [15]. Nevertheless, when the strength of dispersive waves is strong enough, a larger periodic power variation of dispersive waves induced by the interaction between intensity-dependent polarization rotation and passive polarizer [16-19] can be realized by properly setting the linear cavity phase delay. In this paper, we present the experimental observation of modulation instability induced by periodic power variation in a passively mode-locked fiber ring laser. Modulation instability with different wavelength shifts of subsideband generation induced by different periods of power variation was observed.

## 2. Physical mechanism and experimental results

As the sideband is stable on the optical spectrum, the phase delay between soliton pulse and dispersive waves must be multiple of $2\pi$ in a single round trip for the constructive interference of them. Correspondingly, the phase delay between soliton pulse and dispersive wave is fixed at the polarizer. Combining the actions of optical Kerr effect and polarizer, the dispersive waves experience the same period of power fluctuation as the soliton pulse in the laser cavity. Although the dispersive waves experience periodic dispersion variation and its period is equal to cavity length in our fiber laser, the gain peak position of modulation instability is determined by the periodic power variation when the period of power variation is lager than the dispersion variation [20]. Our experimental results suggest that the periodic power variation of dispersive waves is the main origin of modulation instability with subsideband generation.

Since the power variation of dispersive waves caused by the interaction between the nonuniform polarization state and the passive polarizer is perfectly periodic, the phase grating induced by the periodic power variation of dispersive waves in the fiber provides a phase-matching requirement of the four-wave mixing process which lead to the subsideband generation. The modulation instability with subsideband generation occurs when the phase-matching condition is satisfied [14,20]:

$$k_{s+} + k_{s-} = 2k_d + k_g. \tag{1}$$

where, the wave number of the virtual grating $k_g = (2\pi n)/L_a$, $n = 0, \pm 1, \pm 2, \pm 3...$, $L_a$ is the distance of the period of the power variation where dispersive wave returns to its original power in the laser cavity, $k_d = -\gamma \bar{P}_0 + \beta(\omega_d)$ is the wave number of the dispersive wave, where $\bar{P}_0$ is average power of dispersive waves and $\beta(\omega_d)$ is the wave number at the central frequency of dispersive waves. These two wave numbers of subsideband are $k_{s+} = k_{s-} = \frac{1}{2}\beta_2\omega_n^2 + \beta(\omega_d)$. Substituting these relations into Eq. (1), the subsideband frequency shifted from the frequency of the dispersive waves is obtained, which is essentially shifted from the frequency of sidebands:

$$\omega_n = \pm[\frac{1}{\beta_2}(\frac{2\pi n}{L_a} - 2\gamma\bar{P}_0)]^{\frac{1}{2}}. \tag{2}$$

Eq. (2) presents that the modulation instability with subsideband generation induced by the periodic power variation arises both in negative and positive group-velocity dispersion region. From Eq. (2), we can find that the subsideband frequency shift scales as:

$$\omega_n \propto \sqrt{\frac{1}{L_a}}. \tag{3}$$

To demonstrate the effect of the periodic power variation on the modulation instability with subsideband generation in a fiber laser, we constructed an all-fiber ring laser shown in Fig.1. A 5m long erbium-doped fiber with positive group-velocity dispersion and 13.53m single mode fiber (SMF-28) comprise the ring cavity, its fundamental repetition is 11.02 MHz. The absorption coefficients of the erbium-doped fiber are 12.5 dB/m at 980 nm and 20.8 dB/m at 1530 nm, respectively. In order to prevent any perturbation to of fiber birefringence, all the fibers were fastened to the optical table. Unidirectional operation and polarization selectivity are provided by the polarization-dependent isolator. Since the fiber laser is a ring structure, the fiber laser could be described as a length of birefringent fiber with two polarizers at both ends. Due to the interaction between the polarization states of circulating light and polarizer, the passive polarizer in the isolator is regarded as an attenuator of the soliton pulse and dispersive waves which offers the condition of the periodic power variation [21]. Two polarization controllers are used to adjust the polarization state of the pulses in the cavity. The output soliton pulse is taken via a 1% coupler. The periodic /nonperiodic power output of pulses is measured by a radio frequency spectrum analyzer (RFSA), and an optical spectrum analyzer (OSA) is used to measure the soliton spectrum. The soliton pulse duration was also measured by using a commercial autocorrelator, which was about 400 fs.

The nonlinear polarization rotation (NPR) technique was used to achieve the self-started mode locking state of the fiber laser. The mode locking threshold was about 40 mW in our fiber laser. In the experiment, we increased the pump power to the mode locking threshold, by carefully adjusting the polarization controllers, the fiber laser achieved the mode locking state. Keep the orientation of polarization controllers, the mode locking operation could be attained again by increasing the pump power. It is always deemed that the output pulses of a fiber laser have the same intensity and profile, as we called average soliton [22]. However, after the fiber laser achieved the mode locking operation, by properly setting the orientation of polarization controllers, the periodic power fluctuation of soliton pulses appears on the RFSA which indicates that the periodic power variation of dispersive waves occurs.

When the fiber laser operated in mode locking state initially, there was no subsideband on the optical spectrum. For better comparing the modulation instability with subsideband generation, we have shown the optical and RF spectrum without modulation instability. Fig.2 (a) shows a typical optical spectrum without subsideband generation. The 3dB spectral bandwidth of the pulse is 7.56 nm. Fig.2 (b) illustrates that in this case the RF spectrum shows only fundamental cavity round-trip frequency which indicates that the pulse-train was uniform. We selected an appropriate pump power and fixed it, by rotating the orientation of polarization controllers, subsideband appeared on the optical spectrum. If the polarization controllers were not finely set, the sidebands experienced a small power variation when the subsideband appeared in the experiment. Nevertheless, the sideband could be stable again by carefully rotating the polarization controllers. The stability of sideband indicates that the phase delay between soliton pulse and dispersive wave is fixed at the polarizer. As a result, the dispersive waves and the soliton pulse share the same period of power fluctuation in the laser cavity. Different wavelength shifts of subsidebands were observed by simply rotating the polarization controllers. It is found that each wavelength shift of subsideband corresponds to a specific period of power variation. These subsidebands exhibit the power tuning characteristic of modulation instability. When we altered the orientation of the polarization controllers slightly, the intensity of subsideband also changed. For example, with one polarization controller fixed, the rotational range of the other polarization controller is about 20 degrees for the subsideband with 2.04 nm wavelength shifts where could change the intensity of subsideband and approximate 5 degrees for 4.2 nm wavelength shifts. Provided that we keep on rotating one polarization controller towards the same direction while other cavity parameters were fixed, the subsideband disappears and the fiber laser is no longer mode locking. In the experimental observation, we found that the larger wavelength shifts corresponded to a narrower tunable range where could change the intensity of subsideband and the modulation depth of dispersive waves is directly proportional to the intensity of subsideband [19].

A noticeable difference between the optical spectra with and without modulation instability is that two new discrete spectral components, which we called subsideband, appear on both sides of the sideband. Fig.3 (a) shows the spectrum of subsideband generation for 2.04 nm wavelength shifts. We found that the CW excitation always present on the spectrum as this kind of the subsideband generated. The intensity of the subsidebands are asymmetric in the experimental observation. This is probably related to the asymmetric gain spectrum of modulation instability as the power variation period is larger than the dispersion-management period which is equal to the cavity length in our fiber ring laser [20]. It is also to note that the intensity of sidebands is also asymmetric due to the wavelengt-dependent loss caused by the NPR technique. Associated with the appearance of subsidebands, two new frequency components also appeared in the RF spectrum. Fig.3 (b) illustrates the RF spectrum of the output pulses corresponding to the spectral state in Fig.3 (a), it presents not only the fundamental cavity round-trip frequency but also two stable frequency sidebands beside it. The frequency sidebands are symmetric on the both sides of the main peak. This indicates that the power of the soliton pulses began to vary periodically and the periodic power fluctuation of the dispersive waves appeared. From Fig.3 (b) we found that the modulation frequency between the main peak and sideband is $f_m = 3.03$ MHz. Then we obtained $L_a = \frac{f_r}{f_m} L_r = 3.64 L_r$, where $f_r$ and $L_r$ are the fundamental cavity round-trip frequency and the one round trip distance of the laser cavity, respectively. This accounts for the soliton pulse periodically returning to its original intensity after every 3.64 round trips in the cavity which is also the period of the power variation of dispersive waves.

For confirming the dependence of wavelength shifts on the period of power variation, with all other cavity parameters fixed, we carefully tuned the orientation of polarization controllers to obtain the different period of power variation. Fig.4 (a) shows the spectrum of the subsideband with 4.2 nm wavelength shift. The CW excitation was not observed when the subsideband with the large wavelength shifts appeared on the spectrum. Fig.4 (b) shows the RF spectrum corresponds to the optical spectrum as shown in the Fig.4 (a). The frequency between the main peak and sideband is 4.39 MHz, which indicates that the dispersive wave returns to its original value after every 2.51 round trips in the laser cavity. Correspondingly, $L_a = 2.51 L_r$.

In our recent experiments, we also constructed an all-fiber ring laser whose configuration is the same as Fig.1, but the erbium-doped fiber used in this experiment is 4.5 m and the single mode fiber is 13.5 m, the laser output was taken by a 10% coupler. The pulse repetition rate is 11.31 MHz. In this experiment, we also observed the modulation intability with subsideband generation caused by one period of power variation as discussed above. In addition, we found that two periods of power fluctuation could coexist in our fiber laser. Correspondingly, two wavelength shifts of subsideband appeared synchronously on the optical spectrum, as shown in Fig.5 (a). The two wavelength shifts of subsidebands are 1.0 nm and 3.52 nm, respectively. Fig.5 (b) shows the RF spectrum corresponding to the spectral state in Fig.5 (a). Four frequency sidebands appear symmetrically to the main peak: the inner two components have a modulation frequency of 0.86 MHz shift from the main peak, and the outside components have a frequency of 4.86 MHz. Accordingly, the two periods of power variation of dispersive waves are $L_a = 13.22 L_r$ and $L_a = 2.77 L_r$, respectively. By carefully rotating the orientation of polarization controllers, we could easily obtain other different periods of power variation. As a result, different wavelength shifts of subsideband were observed on the optical spectrum. As indicated in Eq. (3), the wavelength shift of subsideband varies with the period of power variation. To further demonstrate the dependence of wavelength shifts on the period of power variation, we show in Fig.6 the experimentally measured wavelength shifts with the periodic distances of power variation. As evidenced in Fig.6, the wavelength shifts of subsidebands are inversely proportional to the periods of the power variation.

## 3. Conclusion

In conclusion, we have experimentally observed the modulation instability with subsideband generation induced by the periodic power variation of dispersive waves in passively mode-locked soliton fiber ring lasers. It is found that the periodic power variation is caused by the interaction between the polarization state of circulating light and the polarizer in the laser cavity. Different wavelength shifts of subsideband induced by different periods of power variation were observed. These kinds of subsideband generation were attained by simply changing the linear cavity phase delay. We also have experimentally demonstrated that the wavelength shift of subsideband has an inverse relation with the period of power variation.

This work was supported by the Natural Science Foundation of Guangdong Province, China (Grant No. 04010397).

**Figure Captions**

Fig.1. Schematic of the fiber ring laser cavity. PC: polarization controller; PD-ISO: polarization-dependent isolator; WDM: wavelength division multiplexer.

Fig.2. A typical output (a) optical spectrum and (b) RF spectrum without modulation instability.

Fig.3. Observation of modulation instability with subsideband generation induced by periodic power variation. (a) Optical spectrum of subsideband with 2.04 nm wavelength shift. (b) RF spectrum output corresponds to the spectral state in (a).

Fig.4. (a) Subsideband with 4.2 nm wavelength shift. (b) RF spectrum corresponds to the optical spectrum in (a).

Fig.5. Optical spectrum (a) and corresponding RF spectrum (b) where two wavelength shifts of subsideband coexist in the fiber laser.

Fig.6. Experimentally measured wavelength shifts vs the periodic distance of power variation.

Fig.1

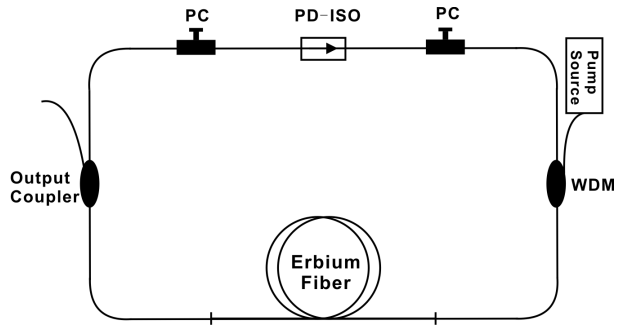

Fig.2

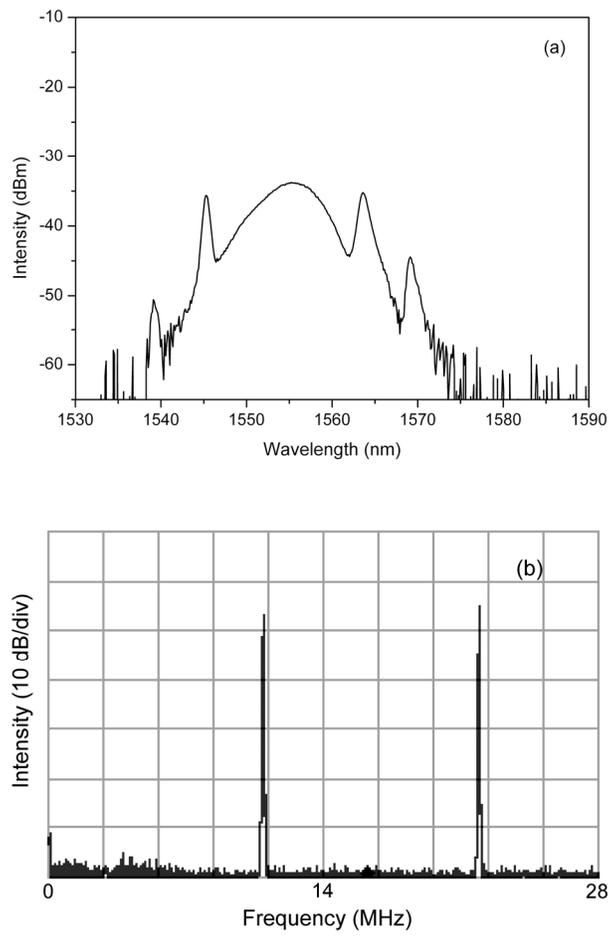

Fig.3

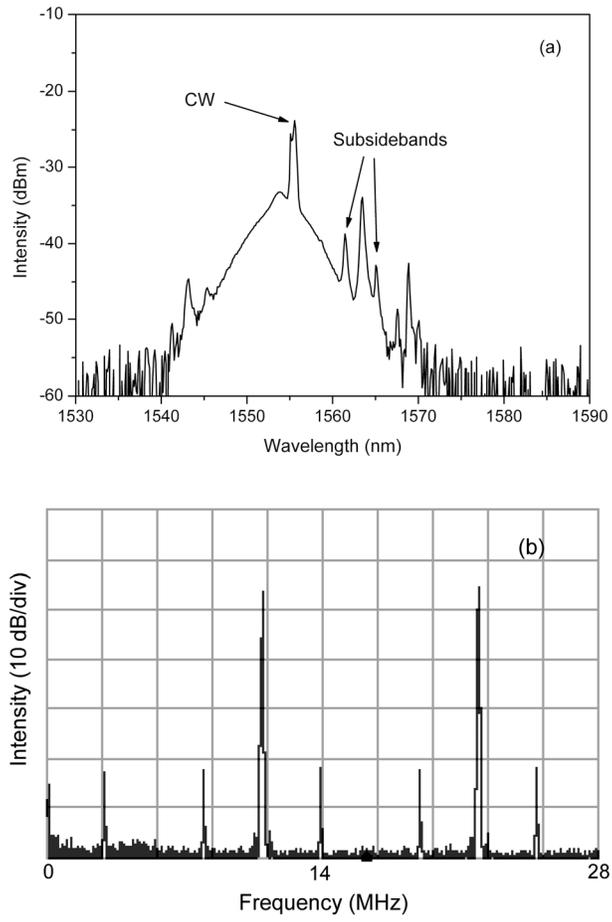

Fig.4

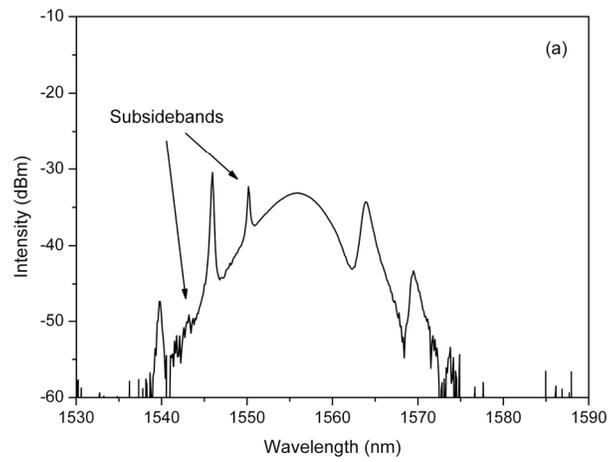

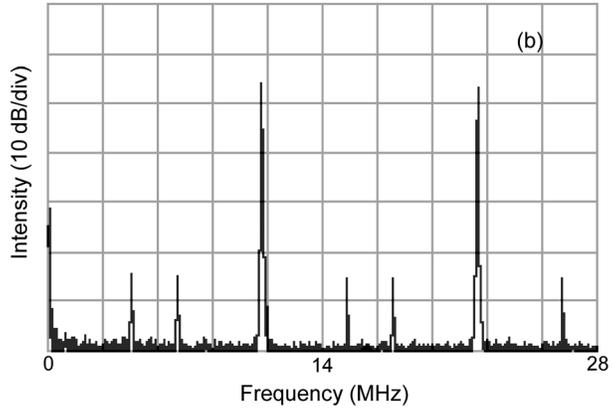

Fig.5

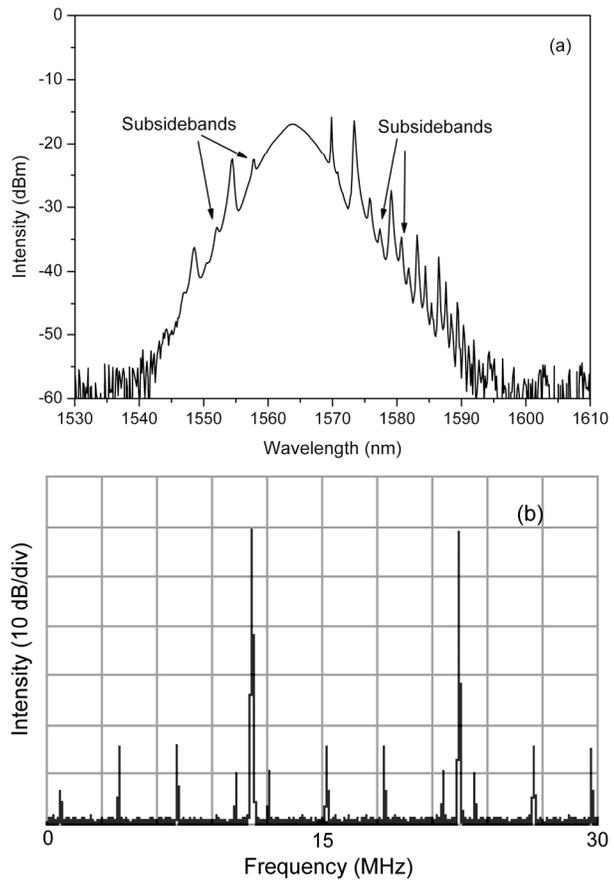

Fig.6

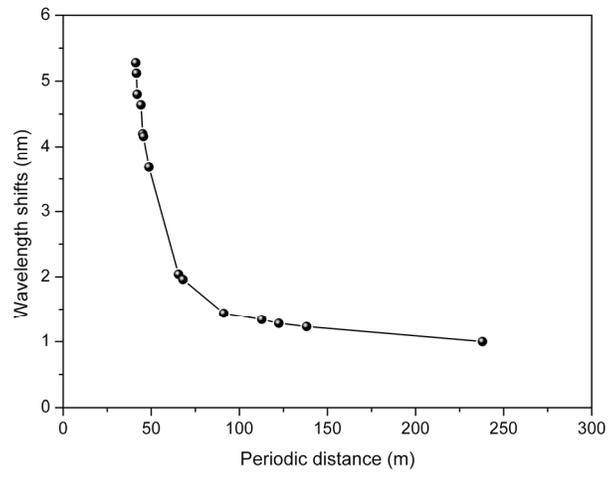